\documentclass[aps,prx,reprint,superscriptaddress,floatfix]{revtex4-2}
\usepackage{amsmath,amssymb,amsfonts,mathtools}
\usepackage{graphicx}
\usepackage{bm}
\usepackage{hyperref}
\hypersetup{hidelinks}
\usepackage{booktabs}
\usepackage{array}
\usepackage{xcolor}
\usepackage{microtype}
\usepackage{placeins}

\newcommand{\vp}{v'}
\newcommand{\Cest}{\mathcal{C}}
\newcommand{\NS}{\mathrm{NS}}
\newcommand{\R}{\mathrm{R}}

\begin{document}



\title{Gravitational anomaly spectroscopy of a Kitaev spin liquid}



\author{Kumar Ghosh}
\email{jb.ghosh@outlook.com}
\affiliation{E.ON Digital Technology, Hannover, Germany}

\begin{abstract}
The chiral central charge $c$ measures the gravitational anomaly of a two-dimensional topological phase and fixes the heat
current along its edge. It has been accessed through thermal transport, where phonons complicate the signal, or through
wave-function probes that need a pure ground state. Here we show that a calibrated combination of directional edge
absorption weights determines $c$. A modulation of the edge couplings that travels along the edge is absorbed according to
the Virasoro algebra, and a static change of the same couplings measures the unknown probe normalization. Two harmonics then
cancel the dependence on the edge state, so the estimator returns $c$ in any state of the ideal edge theory. A ratio of line
weights with and without a vison gives the conformal spin of the Ising anyon. For the Kitaev honeycomb model, exact
calculations recover $c=1/2$ and $h_\sigma=1/16$ to $10^{-5}$ and give $c$ within about $1\%$ on 96 spins. Compiled into gate
circuits on the physical spins of a square-lattice quantum simulator, the protocol returns $c$ within $5\%$ on 48 qubits.
For stochastic gate errors it requires two-qubit error rates of $10^{-6}$ to $10^{-5}$.
\end{abstract}

\maketitle


A two-dimensional topological phase with broken time-reversal symmetry carries chiral edge modes. They move energy in one
direction along the boundary. The amount is fixed by a single number, the chiral central charge $c$: the edge carries the
thermal Hall conductance $\kappa_{xy}/T=c\,\pi^2k_B^2/3h$~\cite{Kane1997}. In field theory $c$ is the coefficient of the
gravitational anomaly~\cite{AlvarezGaume1984}. The edge theory alone cannot conserve energy and momentum when it is placed
in a curved space, and the bulk compensates for it. For the non-Abelian phase of the Kitaev honeycomb model~\cite{Kitaev2006}
$c=\frac12$. Its edge is a single chiral Majorana mode, and its bulk hosts Ising anyons $\sigma$ with conformal spin
$h_\sigma=\frac1{16}$, the excitations proposed for topological quantum computation~\cite{Nayak2008}.

Measuring $c$ is hard. Thermal Hall experiments have reported half-integer values in the $\nu=\frac52$ quantum Hall
state~\cite{Banerjee2018}, where the conductance gives the net central charge only once counter-propagating edge modes have
equilibrated, and in the Kitaev material $\alpha$-RuCl$_3$~\cite{Kasahara2018}. In $\alpha$-RuCl$_3$ the interpretation has
been debated, partly because phonons also produce thermal Hall signals; geometry-dependent measurements now support an edge
origin~\cite{Zhang2025}. Ground-state wave functions give $c$ and topological spins through momentum
polarization~\cite{Tu2013}, and the modular commutator gives $c$ from the reduced density matrices of three adjacent
regions~\cite{KimShi2022}. These routes require a pure ground state and tomography.

Programmable quantum processors now prepare the Kitaev spin liquid. Chiral edge propagation was observed on 22 trapped-ion
spins~\cite{Ali2025}, and an odd Chern number of the free-fermion state was verified on neutral atoms~\cite{Evered2025}. The
central charge of one-dimensional critical chains has been measured on a processor from entropies of Pauli
measurements~\cite{IBMc2024}, and conformal spectra of such chains have been resolved on a Rydberg simulator~\cite{Sun2026}.
In those spectra the central charge enters only as an overall energy shift, which was not resolved. These chains are also
non-chiral: their left and right movers carry the same central charge, so their gravitational anomaly vanishes. To our
knowledge, the chiral central charge of a two-dimensional phase has not been measured on any processor. Can $c$ be read off
from a measurement that such devices can perform, without a pure ground state?

\begin{figure*}[htb]
\centering
\includegraphics[width=\textwidth]{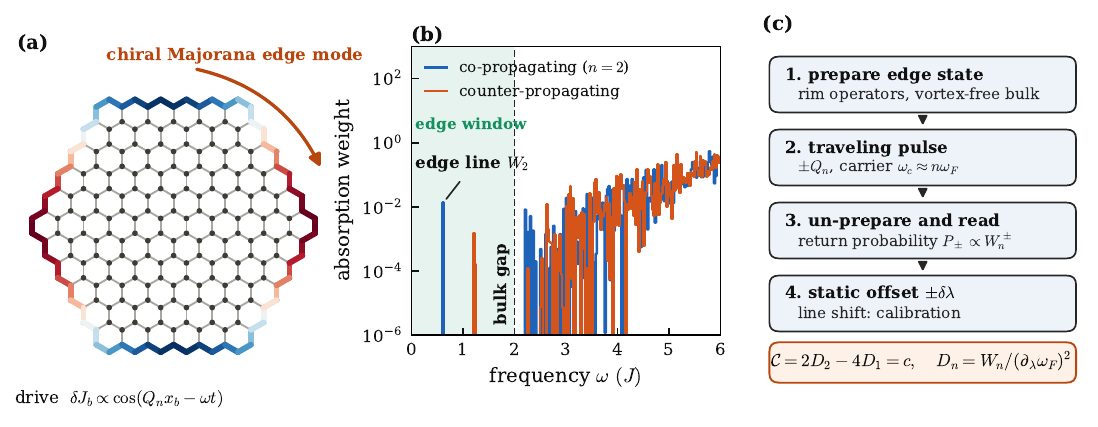}
\caption{Principle. (a) Hexagonal flake of the Kitaev honeycomb model (216 spins). The edge couplings are modulated by a
traveling wave $\delta J_b\propto\cos(Q_nx_b-\omega t)$; the colors show a snapshot for $n=2$. The edge hosts one chiral
Majorana mode. (b) Exact absorption spectra of the two propagation directions on a 384-spin flake ($\kappa=0.3$). The drive
that co-propagates with the edge mode has a strong line below the bulk gap, and the counter-propagating drive has only a much
weaker subgap line. The line weight $W_2$ is the input of the sum rule. The shaded region is the edge window, and the bulk
continuum above the gap is common to both directions. (c) The echo protocol for a pure initial state. In step 3, $P_\pm$ is the
probability that the pulse has excited the system, one minus the return probability. Rim operators, or a direct circuit
preparation (Fig.~\ref{fig:qsim}), fix the unpaired boundary Majoranas.}
\label{fig:concept}
\end{figure*}

Here we show that it can. The edge stress tensor, the operator that measures energy density along the edge, generates the
Virasoro algebra, and $c$ is its central term. In a conformal tower $c$ cancels from every excitation gap, but it survives in
the transition weights of the stress tensor. A traveling modulation of the edge couplings couples to the stress tensor
[Fig.~\ref{fig:concept}(a)], and its absorption measures a Virasoro commutator times an unknown probe coupling. A static
change of the same couplings measures that coupling, with no value of $c$ assumed. A combination of two harmonics then
removes the only state-dependent term and returns $c$ in any edge state. A second combination, the ratio of line weights with
and without an anyon in the bulk, needs no calibration and yields $h_\sigma$.

Spectral weights of chiral edges have been studied before. Local Raman probes of Kitaev edges give a weight that grows as
$\omega^3$~\cite{HeNagaosa2021}. This is the central term $cn^3/12$ of the Virasoro norm at large $n$, with a prefactor that
contains the unknown coupling of light to the spins. Inelastic light scattering was proposed as a signature of the
gravitational anomaly in the bulk of fractional quantum Hall states~\cite{Wiegmann2018}. Circular dichroism of drives on a
quantum Hall edge measures the many-body Chern number~\cite{Unal2025}. There the density drive projects onto U(1) current
modes, whose commutator is a number set by the Hall conductance. Our drive is the gravitational counterpart: it projects onto
Virasoro generators, whose commutator contains $c$ and also $L_0$, which depends on the edge state. A neutral Majorana edge has
no charge density to drive, and its infrared absorption is gapped~\cite{HeNagaosa2021}, so the stress tensor is its natural
channel. Most recently, the Virasoro transition norms $2nh+\frac{c}{12}(n^3-n)$ were derived for shape deformations of
Moore--Read interfaces, together with sector ratios in which the unknown coupling cancels~\cite{Ma2026}. Our contribution is
the step from spectral weights to an absolute and state-independent number, its microscopic test in a lattice spin liquid,
and its translation into quantum circuits.

We establish the route for the Kitaev model with exact calculations on up to 7168 spins. Calibrating with the shift of the
measured line strongly reduces finite-size errors in the systems we tested, so that about 100 spins suffice. The result
survives thermal and non-stationary edge states, static vison disorder, random coupling errors and interactions between the
edge Majoranas in a fixed flux sector. It fails when bulk Majorana quasiparticles are present, which sets a clear experimental
requirement. Finally, we compile the protocol into gate circuits on the physical spins of a square-lattice quantum simulator,
verify these circuits with matrix-product-state simulations, and derive the error rates that the measurement requires.

\section*{A calibrated sum rule for chiral edges}\label{sec:principle}

The low-energy theory of a chiral edge of length $L$ is a chiral conformal field theory (CFT) with velocity $v$. Its
Hamiltonian is $H_{\rm edge}=\omega_F\,(L_0-\frac{c}{24})$, with level spacing $\omega_F=2\pi v/L$~\cite{DiFrancesco1997}.
The operators $L_n$ are the Fourier modes of the edge stress tensor. They obey the Virasoro algebra,
\begin{equation}
[L_m,L_n]=(m-n)L_{m+n}+\frac{c}{12}\,m(m^2-1)\,\delta_{m+n,0},
\label{eq:virasoro}
\end{equation}
with $L_n^\dagger=L_{-n}$. $L_{-n}$ with $n>0$ creates an excitation of momentum $2\pi n/L$ and energy $n\omega_F$ moving in
the chiral direction; $L_n$ removes one. The central charge is the last term. In a conformal tower the energies are
$\omega_F(h+N-c/24)$, so $c$ cancels from every excitation gap. It survives in the transition weights of the stress tensor,
and these are what an absorption measurement records. For the Kitaev edge the CFT is the chiral Ising theory with
$c=\frac12$~\cite{Kitaev2006}.

To reach these weights we modulate the couplings $o_b$ of the edge bonds with a traveling phase, a drive $\lambda O_n$ with
\begin{equation}
O_n=\sum_b e^{iQ_nx_b}\,o_b,\qquad Q_n=\frac{2\pi n}{L},
\label{eq:drive}
\end{equation}
where $x_b$ is the position of bond $b$ along the edge. At low energy a bond coupling acts on the edge as a constant plus the
stress tensor at $x_b$. The sum over bonds therefore selects one Fourier mode, $O_n\to a(2\pi/L)L_{-n}$, as in the lattice
construction of Virasoro generators from Fourier modes of the Hamiltonian density~\cite{KooSaleur1994}. The coefficient $a$
is a microscopic property of the probe and is not known. We collect the absorbed weight $W_n$ of all transitions below the
bulk gap and subtract the weight of the counter-propagating drive $O_{-n}$. This removes contributions that are nearly the same
for both directions, mainly from the bulk. Fermi's golden rule then gives $W_n=a^2(2\pi/L)^2\langle[L_n,L_{-n}]\rangle$, in
whatever state the edge is: pure or mixed, stationary or not (Methods and Supplementary Note~\ref{si:sumrule}).

A static change $\lambda O_0$ of the same couplings measures the prefactor. It shifts the level spacing at the rate
$\partial\omega_F/\partial\lambda=a\,2\pi/L$. This rate follows from first-order perturbation theory: the energy of each level
shifts by the expectation value of $O_0$ in that level (the Hellmann--Feynman theorem~\cite{Feynman1939}). Dividing $W_n$ by
its square defines the calibrated weight
\begin{equation}
D_n\equiv\frac{W_n}{(\partial\omega_F/\partial\lambda)^2}=2n\langle L_0\rangle+\frac{c}{12}(n^3-n).
\label{eq:Dn}
\end{equation}
Equation~\eqref{eq:Dn} contains one unknown property of the state, $\langle L_0\rangle$, and one universal number, $c$. Two
harmonics separate them:
\begin{equation}
\Cest\equiv2D_2-4D_1=2\Big(4\langle L_0\rangle+\frac{c}{2}\Big)-4\big(2\langle L_0\rangle\big)=c .
\label{eq:C}
\end{equation}
The calibration removes the probe coupling, and the harmonic combination removes the edge-state contribution. Neither step
assumes the value of $c$. Equation~\eqref{eq:C} holds in any state of the ideal edge theory.

The same weights also measure the anyon. Without a vison in the bulk, the edge Majorana is antiperiodic around the edge
(Neveu--Schwarz, NS sector), and the edge ground state has $L_0=0$. Then $D_1=0$ and $D_2=c/2$. A vison inside the edge contour
carries a $\pi$ flux and makes the edge Majorana periodic (Ramond, R sector). The edge ground state then has
$L_0=h_\sigma=\frac1{16}$~\cite{DiFrancesco1997,Kitaev2006}, and $D_1=2h_\sigma$. Since the prefactor $a$ is the same in both
sectors, the ratio of the bare line weights needs no calibration:
\begin{equation}
\Lambda\equiv\frac{W_1^\R}{W_2^\NS}=\frac{2h_\sigma}{c/2}=\frac{4h_\sigma}{c}=\frac12 .
\label{eq:Lambda}
\end{equation}
Unlike $\Cest$, the ratio $\Lambda$ refers to the two sector ground states, measured with the same geometry and drive. It
removes the common probe normalization, but the relative drive amplitudes and readout efficiencies of the two measurements
must still be controlled. Together with $\Cest$ it gives $h_\sigma$. Within the Ising description this fixes the topological
spin $e^{2\pi ih_\sigma}$ of the anyon; it does not measure braiding.

In practice the calibration can be taken in two ways. The \emph{ladder} calibration takes $\partial\omega_F/\partial\lambda$
from the shift of the fitted ladder of edge levels. The \emph{line} calibration takes it from the shift of the measured
absorption line itself, $\partial\omega_{\rm line}/\partial\lambda=n\,\partial\omega_F/\partial\lambda$. This is what a pulse
experiment measures. In both versions the calibration is a measured level shift (Methods).


\section*{Central charge and anyon spin of the Kitaev edge}\label{sec:kitaev}

We test the sum rule in the Kitaev honeycomb model. Spins $\frac12$ sit on the sites of a honeycomb lattice. Each bond couples
one spin component, $\sigma^x$, $\sigma^y$ or $\sigma^z$, set by the bond direction~\cite{Kitaev2006}:
\begin{equation}
H=-J\sum_{\langle jk\rangle_\alpha}\sigma^\alpha_j\sigma^\alpha_k-\kappa\sum_{\langle jkl\rangle}\sigma^\alpha_j\sigma^\gamma_k\sigma^\beta_l .
\label{eq:H}
\end{equation}
The three-spin term $\kappa$ is generated at third order by a magnetic field and opens a gap in the bulk~\cite{Kitaev2006}.
The model maps exactly onto free Majorana fermions in a static $\mathbb Z_2$ gauge field. The ground state has no
flux~\cite{Lieb1994}, and a plaquette with flux $\pi$ is a vison, which binds a Majorana zero mode and is the Ising anyon
$\sigma$. All states we study, including thermal states and the states produced by the drives, are Gaussian states of the
matter Majoranas. All absorption weights therefore follow exactly from their two-point correlations (Methods). We use zigzag
cylinders of up to 7168 spins and hexagonal flakes of up to 3456 spins, and three drives: modulated edge couplings, couplings
with a smooth strain-like profile, and a modulated three-spin term (Methods).

\begin{figure*}[htb]
\centering
\includegraphics[width=\textwidth]{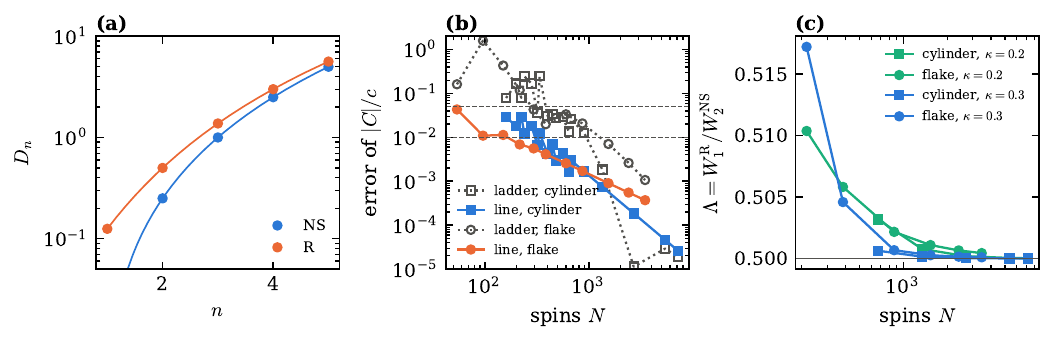}
\caption{The sum rule. (a) Calibrated weights $D_n$ on the $L=512$ cylinder ($\kappa=0.3$, edge drive) without a vison (NS)
and with a vison (R); lines are Eq.~\eqref{eq:Dn} with $c=\frac12$, $h_\sigma=\frac1{16}$. The NS weight at $n=1$ vanishes in
the ideal theory and lies below the logarithmic axis. (b) Error $\big||\Cest|/c-1\big|$ against the number of spins (NS,
$\kappa=0.3$) with the ladder (open) and line (filled) calibration; dashed lines mark $5\%$ and $1\%$. (c) Calibration-free
sector ratio $\Lambda$, Eq.~\eqref{eq:Lambda}.}
\label{fig:theorem}
\end{figure*}

Figure~\ref{fig:theorem}(a) shows Eq.~\eqref{eq:Dn} on the largest cylinders. With the ladder calibration, $\Cest=0.49999$
(NS) and $0.49997$ (R) at $L=512$ and $\kappa=0.3$, and $D_1^\R/2=0.06250$ (Extended Data Table~\ref{tab:largest}). The three
drives differ in $\vp/v$ by up to a factor of 3, yet they agree within $0.03\%$. The approach to the limit goes as $1/L^2$.
Flakes converge in the same way, with $\Cest\to-\frac12$. The sign is the chirality relative to the counter-clockwise
perimeter coordinate. Further sizes and couplings are given in Supplementary Note~\ref{si:bench}.

The line calibration is the practical advance [Fig.~\ref{fig:theorem}(b)]. With the ladder calibration, flakes need about 1500
spins for $1\%$, and the error is erratic below a few hundred spins. On the 96-spin flake the ladder calibration gives
$|\Cest|=1.294$. The line calibration of the same flake gives $0.5055$, within $1.1\%$ of $c$. The line calibration involves
the same lattice states as the line weight, so their common lattice corrections largely cancel in the ratio. This is a
numerical observation for the systems we tested, not a general theorem. The sector ratio $\Lambda$
[Fig.~\ref{fig:theorem}(c)] converges to $\frac12$ at the same rate: it is $0.5046$ at 384 spins and $0.5001$ on the largest
cylinders.


\section*{Robustness and its limits}\label{sec:robust}

The sum rule is an identity of the ideal edge theory. The lattice tests ask where it holds in practice: in which edge states,
under which perturbations and with which bulk.

\begin{figure*}[htb]
\centering
\includegraphics[width=\textwidth]{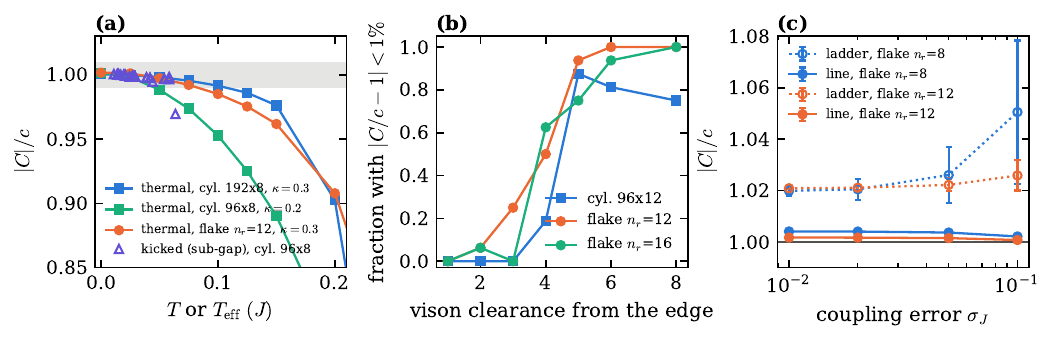}
\caption{Robustness. (a) $|\Cest|/c$ in thermal edge states (calibration in the ground state) and in kicked non-stationary
states, against $T$ or the temperature $T_{\rm eff}$ with the same injected edge energy; the gray band marks a $\pm1\%$
tolerance, not a confidence interval. (b) Static vison gas at density $0.05$ per plaquette: fraction of 16 random
configurations with $\Cest$ within $1\%$, against the minimum distance of any vison from the edge. (c) Random coupling errors of
relative size $\sigma_J$ (24 random samples; flakes, $\kappa=0.3$): ladder against line calibration.}
\label{fig:states}
\end{figure*}

\paragraph*{Edge states.} We take the calibration once, in the ground state, as an experiment would. In thermal edge states
$\Cest$ stays within $1\%$ up to $T=0.1$ on $\kappa=0.3$ cylinders and up to $T=0.05$ to $0.075$ on flakes, while $D_1$ and
$D_2$ grow by large factors [Fig.~\ref{fig:states}(a)]. The deviation at higher $T$ comes from the curvature of the edge
dispersion, which the CFT neglects. The effective temperature reported for the trapped-ion state, $T_{\rm eff}\approx0.2J$~\cite{Ali2025},
lies outside this window. State preparation must therefore reach about half of it. Slow traveling kicks that leave the bulk
unexcited produce non-stationary edge states. On a $96\times8$ cylinder, all such states below the thermal $1\%$ threshold give
$\Cest$ within $0.14\%$ of $c$, also when the probe is delayed. Sudden two-qubit gates on edge bonds behave differently. They create coherences between edge and bulk modes and shift
$\Cest$ by $4\%$ or more, so kicks must be slow on the scale of the bulk gap (Supplementary Note~\ref{si:states}).

\paragraph*{Visons and coupling errors.} Random static visons [Fig.~\ref{fig:states}(b)] leave $\Cest$ unchanged once no vison
is within about five lattice spacings of the edge. On a flake an odd number of visons switches the edge to the R sector.
$D_1/2$ then jumps to $\frac1{16}$ while $\Cest$ stays fixed, as Eq.~\eqref{eq:C} predicts. A processor never sets every
coupling exactly. Independent random errors on the couplings bias the ladder calibration, by $3\%$ on the 384-spin flake at
$\sigma_J=10\%$ [Fig.~\ref{fig:states}(c)]. The line calibration measures the same disordered couplings that it drives. It keeps
$\Cest$ within $0.5\%$ of $c$ up to $\sigma_J=20\%$ on the 864-spin flake.

\paragraph*{Bulk quasiparticles.} A single occupied bulk Majorana mode spoils the measurement on all sizes studied.
Transitions within the bulk band then fall below the gap and into the edge windows. The protocol therefore requires the bulk
Majorana vacuum. Thermal bulk excitations are negligible below $T\approx0.2$; excitations created by errors are the relevant
threat.

\paragraph*{Interacting edge Majoranas.} So far the edge Majoranas were free. Away from the solvable point they interact. A
Heisenberg coupling or a general magnetic field creates visons. Below the vison gap these virtual processes leave a static
gauge field. Their effect on the matter Majoranas is then captured by renormalized quadratic couplings and by four-Majorana
interactions, the lowest-order interactions allowed by the $\mathbb Z_2$ gauge structure. We add a representative interaction
of strength $g$ that couples the two parallel bonds of every plaquette and keep the flux sector fixed (Methods). We find the
lowest states with the density-matrix renormalization group (DMRG)~\cite{White1992}, a variational method over
matrix-product states whose accuracy is controlled by the bond dimension $\chi$. At $g=0$ it reproduces the exact $\Cest$ to
$1.4\times10^{-5}$ on 96 spins. On 150 spins $\chi=256$ is not converged, so we show that scan as a trend only (Extended Data
Table~\ref{tab:dmrg}).

\begin{figure*}[htb]
\centering
\includegraphics[width=\textwidth]{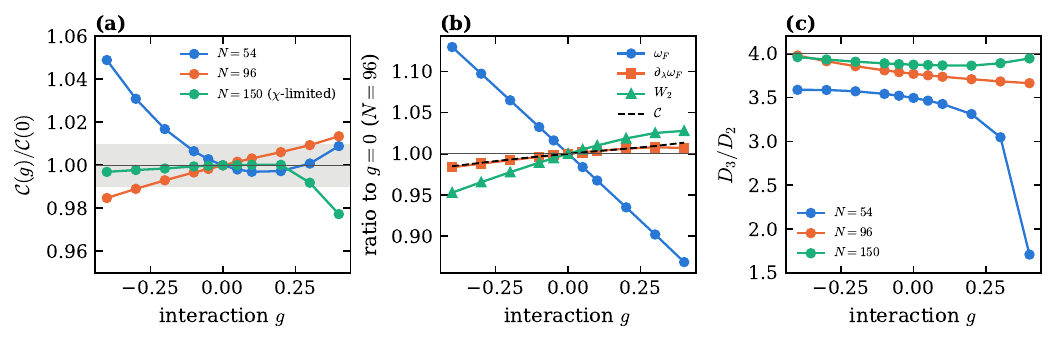}
\caption{Interacting edge Majoranas in a fixed flux sector (DMRG on flakes, $\kappa=0.3$, $\chi=256$). (a) $\Cest(g)/\Cest(0)$,
relative to the free value of each flake; the gray band marks $\pm1\%$. The 150-spin scan is limited by $\chi$ (Extended Data
Table~\ref{tab:dmrg}) in all panels. (b) On 96 spins: level spacing $\omega_F$, calibration rate $\partial_\lambda\omega_F$
and raw line weight $W_2$ relative to $g=0$, with $\Cest$ (dashed). (c) Internal check $D_3/D_2$; the conformal value is 4.}
\label{fig:interactions}
\end{figure*}

The interaction moves the edge strongly [Fig.~\ref{fig:interactions}(b)]. On 96 spins, $\omega_F$ changes from $+13\%$ to
$-13\%$ of its free value. The calibration follows the line weight. As a result, $\Cest$ stays within $1.5\%$ of its free
value over the whole scan, $|g|\le0.4$ [Fig.~\ref{fig:interactions}(a)]. The internal check $D_3/D_2$ rises with size towards
the conformal value 4 and does not depend on $g$ [Fig.~\ref{fig:interactions}(c)]. Within the fixed flux sector and the tested
range, the calibration thus absorbs the renormalization of both the velocity and the drive (Supplementary
Note~\ref{si:interact}).


\section*{From edge spectroscopy to spin circuits}\label{sec:protocol}

We now ask what a quantum simulator needs to carry out the measurement. We first simulate each step exactly in the
free-Majorana sector and then compile the protocol into gate circuits on the physical spins of a square-lattice processor.

\paragraph*{Two readouts.} The correlator readout measures the two-time correlator $F(t)=\langle[O_n^\dagger(t),O_n]\rangle$
directly, as in Ref.~\cite{Ali2025}, and integrates it against a spectral window below the gap. It applies to mixed states.
The drive's weight in the bulk exceeds the edge line by three orders of magnitude [Fig.~\ref{fig:concept}(b)], so the window
must reject the bulk sharply. A $5\%$ result requires a maximum evolution time $t_{\max}=15$ to $30/J$ [Extended Data
Fig.~\ref{fig:protocol}(b)]. The echo readout is a pure-state shortcut [Fig.~\ref{fig:concept}(c)]. The processor prepares the
edge ground state and applies a traveling Gaussian pulse of the edge couplings with its carrier at the line. It then runs the
preparation backwards and measures the probability $P_\pm$ that the pulse has excited the system, for the two propagation
directions. To lowest order $P_\pm=2\pi\sigma_t^2\lambda^2W^\pm_n$, with $\sigma_t$ the pulse width (Methods). The calibration
comes from the same readout, from the line centre at static offsets of the edge couplings. On flakes of 96 to 600 spins the
echo returns $c$ within $3.5\%$ (Extended Data Table~\ref{tab:echo}). For a mixed state the return probability is not the net
absorption, and the correlator readout should be used instead.

\paragraph*{Boundary and Floquet evolution.} Two further requirements follow from the microscopic structure. First, the
unpaired gauge Majoranas of the boundary must not couple to the edge. Reference~\cite{Ali2025} removes their degeneracy with a
weak boundary field. This field couples them to the edge and destroys the signal on cylinders and the R-sector signal on
flakes. Rim operators lift the degeneracy without coupling to the edge (Supplementary Note~\ref{si:floquet}). Second, digital
processors do not evolve with $H$ directly. They apply a periodic sequence of gates, a Floquet circuit, whose evolution over
one period equals evolution with an effective Hamiltonian. The six-layer bond circuit of the trapped-ion implementation obeys
Eq.~\eqref{eq:C} through its exact Floquet Hamiltonian, with $\Cest=0.5004$ on a $192\times8$ cylinder [Extended Data
Fig.~\ref{fig:protocol}(c)].

\begin{figure*}[htb]
\centering
\includegraphics[width=0.9\textwidth]{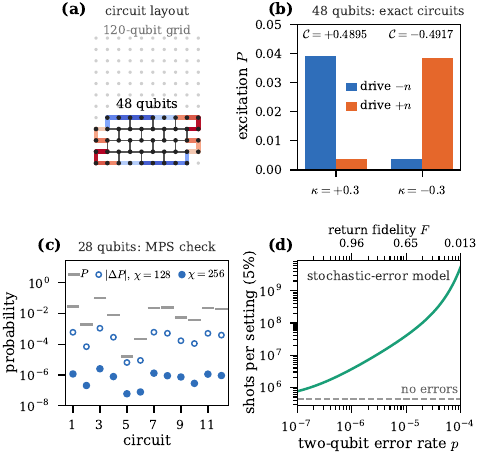}
\caption{Quantum-simulator circuits. (a) Circuit layout: 48-qubit brick-wall patch on a 120-qubit square grid; every bond is a
coupler. Thick edge bonds carry the drive; their colors show a snapshot of $\delta J_b$ for $n=2$. (b) Exact simulation of the
48-qubit circuits: excitation probabilities of the $n=2$ echo ($\lambda=0.1$) for the two drive directions, and the resulting
$\Cest$, whose sign gives the absorbing direction. (c) Independent matrix-product-state (MPS) simulation of the twelve circuits
on 28 qubits, in the order of Methods: error $|\Delta P|$ of the excitation probability at two bond dimensions; gray dashes give
$P$ itself. (d) Stochastic-error model: shots per setting for a $5\%$ standard error on $\Cest$ against the probability $p$ of an
error per compiled two-qubit gate, with a common return fidelity $F=(1-p)^{4.3\times10^4}$ for all circuits and equal shots for
all settings, including the three $\lambda=0$ references. Panels (b) to (d) are simulations; no processor was run.}
\label{fig:qsim}
\end{figure*}

\paragraph*{Gate circuits on a square-lattice processor.} The honeycomb lattice is a brick-wall subgraph of the square lattice.
On a square-lattice processor every spin is then a qubit and every Kitaev bond a coupler [Fig.~\ref{fig:qsim}(a)]. We compile
the echo readout into gate circuits on these physical spins, for a 48-qubit patch of a 120-qubit processor (a calibration
snapshot of IBM Phoenix). Each circuit prepares the flux-free sector with stabilizer gadgets, short Clifford circuits around
each plaquette. It then prepares the Majorana vacuum of one Trotter step with bond gates that act as Givens
rotations~\cite{Langer2026}. It applies Trotter steps of Eq.~\eqref{eq:H} with the pulse in the angles of the edge-bond gates,
undoes the preparation and measures the return to $|0\cdots0\rangle$. No boundary terms are needed. The unpaired gauge
Majoranas never enter a gate, and the preparation fixes them in a state that the gates conserve. The preparation uses the
solvability of the model; away from the solvable point the vacuum would be prepared adiabatically or variationally.

Bond and three-spin gates commute with every plaquette flux and are quadratic in the matter Majoranas. An exact Majorana
simulation therefore follows the circuits gate by gate at any size. On 16 qubits it agrees with a statevector simulation to
$10^{-9}$. On the 48-qubit patch the twelve circuits give $\Cest=0.489$, and $0.490\pm0.016$ with $10^6$ shots per setting.
Reversing $\kappa$ moves the absorption to the other drive direction and gives $\Cest=-0.492$ [Fig.~\ref{fig:qsim}(b)].
Varying the pulse width and its truncation moves $\Cest$ between $0.48$ and $0.52$, a finite-pulse systematic of the estimator.
As a check that does not rely on the Majorana structure, we simulate the full spin state with the matrix-product-state
simulator of Qiskit Aer~\cite{Vidal2003,Qiskit2024}. The $\mathbb Z_2$ gauge sector adds a flat entanglement spectrum across
every cut~\cite{YaoQi2010}, and the 48-qubit circuits would need a bond dimension $\chi\approx10^3$. On a 28-qubit patch,
twelve circuits of the same construction agree with the exact excitation probabilities to $3\times10^{-6}$ at $\chi=256$
[Fig.~\ref{fig:qsim}(c)]. This patch is too narrow to give $c$, but it tests every block of the circuits.

\paragraph*{Errors and resources.} The vacuum preparation and the pulse commute with every plaquette flux and with the pairing
of the boundary Majoranas. A single-qubit Pauli error in these parts anticommutes with one of them, so the state cannot return
to $|0\cdots0\rangle$. Such errors lower the return fidelity $F$ of all circuits alike, and three $\lambda=0$ reference
circuits measure it. The remaining cost is statistical. A circuit compiles to about $4.3\times10^4$ two-qubit gates. For
stochastic errors at a rate $p$ per gate, a $5\%$ error on $\Cest$ then needs $4\times10^6$ shots per setting at $p=10^{-6}$
and $4\times10^7$ at $p=10^{-5}$, against $4\times10^5$ without errors [Fig.~\ref{fig:qsim}(d)]. Coherent and
setting-dependent errors are not removed by the reference (Methods). Present physical error rates, $p\sim10^{-3}$, leave
$F\sim10^{-19}$.

\paragraph*{Resources.} Both readouts need coherent evolution for about $10$ to $25/J$ on about 50 to 200 spins, with the bulk in its
Majorana vacuum, no boundary field, and $10^5$ to $10^6$ shots per setting in the absence of errors. This exceeds the depth of
the present 22-spin experiments~\cite{Ali2025}. For stochastic errors the measurement requires two-qubit error rates of
$10^{-6}$ to $10^{-5}$, which makes it a natural benchmark for early fault-tolerant processors. The calibration-free ratio $\Lambda$ is the
natural first target, since it needs no static offsets.


\section*{Scope and outlook}\label{sec:outlook}

The directional absorption of a traveling edge drive, normalized by a static calibration of the same couplings, measures the
central term of the Virasoro algebra. Three features distinguish this route from existing probes. First, it does not need a
pure state or tomography. The sum rule~\eqref{eq:C} is an operator identity of the edge theory, so it holds in thermal and
non-stationary edge states, which a processor or a material naturally provides. The correlator readout inherits this
property; the echo readout assumes a pure initial state. Second, it isolates the edge by its chirality and its frequency. This
separates the edge response from bulk contributions, which in thermal transport include phonons. It also needs no heat flow,
so it does not rely on equilibration between edge modes or between the edge and phonons~\cite{Banerjee2018,Zhang2025}. Third,
the same measurement yields the conformal spin of the non-Abelian anyon through the calibration-free ratio $\Lambda$, and,
within the Ising description, its topological spin. The measurement is insensitive to static visons away from the edge, to
random coupling errors and to interactions between the edge Majoranas. It requires the bulk Majorana vacuum and pulses that are
slow on the scale of the bulk gap.

The sum rule uses only the Virasoro algebra of the edge. It therefore applies to any chiral edge described by a CFT, such as
quantum Hall edges and the edges of chiral superconductors, provided a drive couples to the edge stress tensor and a static
change of the same coupling is available. In Kitaev materials, traveling strain waves along a sample edge could play the role
of the drive, with a static strain as the calibration. The Raman vertex of the Kitaev model is also a sum of bond
energies~\cite{HeNagaosa2021}, so inelastic light scattering at an edge couples to the same operator, although its calibration
would need a static perturbation with the same coupling. Edges with several branches, such as those proposed for
$\nu=\frac52$, are an open case. Branches with different velocities give separate lines, so directional spectroscopy might
resolve an upstream neutral mode that heat transport sees only after equilibration~\cite{Banerjee2018}.

For quantum simulation, the measurement tests the predicted anomaly and anyon data of a prepared state with a definite number
and a definite resource cost. Our exact results are for the solvable model, and the DMRG results extend them to interacting
edge Majoranas in a fixed flux sector. The remaining step toward the full Kitaev--Heisenberg and field problem is a controlled
treatment of the gauge sector at sizes where the edge spectrum lies below the vison gap. Within the non-Abelian phase $\Cest$
should stay at $c$. A change of $\Cest$ could then mark the boundary of the phase, once loss of spectral isolation, calibration
failure and preparation errors are excluded.


\section*{Methods}

\subsection*{Kitaev model, flux sectors and boundaries}

We set $J=1$ in Eq.~\eqref{eq:H} and use $\kappa\in\{0.1,0.2,0.3\}$. The bulk gaps are $1.04$, $2.0$ and $2.0$; for
$\kappa\ge0.2$ the gap saturates at $2J$. The edge velocities are $1.11$, $1.87$ and $2.30$. Kitaev's solution writes each spin
with four Majorana fermions, $b^x,b^y,b^z$ and $c$, which are real fermion operators ($c^\dagger=c$,
$c^2=1$)~\cite{Kitaev2006}. The bond variables $u_{jk}=ib^\alpha_jb^\alpha_k=\pm1$ are conserved and form a static
$\mathbb Z_2$ gauge field. For fixed $u$, $H=\frac{i}{4}\sum_{jk}A_{jk}c_jc_k$ describes free $c$ Majoranas. The
gauge-invariant flux through a hexagon is the plaquette operator $W_p=\pm1$. The ground state has no flux ($W_p=1$
everywhere)~\cite{Lieb1994}; a plaquette with $W_p=-1$ is a vison. In the non-Abelian phase each vison binds a Majorana zero
mode and is the Ising anyon $\sigma$. States of free fermions, including thermal states and the states produced by the drives,
are Gaussian states: they are fully specified by the two-point correlation matrix $G_{jk}=\langle c_jc_k\rangle$. All
absorption weights follow exactly from $G$ (see Gaussian states below).

An open edge leaves one gauge Majorana $b$ unpaired on every outer site. We pair them with boundary operators
$R=\sigma^x_i\sigma^z_k\sigma^y_j$, products of spin components along the edge chosen to commute with every term of
Eq.~\eqref{eq:H}. In the Majorana picture they act only on the unpaired $b$'s. Exact diagonalization of a 16-spin cylinder
confirms $[R,H_K]=[R,H_\kappa]=0$ to machine precision and that a term $-h_R\sum R$ removes the degeneracy of the unpaired
modes, leaving the $c$-Majorana edge unchanged. We call these the rim operators.

We use zigzag cylinders ($L\le512$, width $W\le12$) and hexagonal flakes ($N=6n_r^2\le3456$ spins), built with
Kwant~\cite{Kwant}. The R sector is a flux through the cylinder or a vison at the flake centre. On flakes $x_b$ in
Eq.~\eqref{eq:drive} is the arc length along the perimeter, and the sign of $\Cest$ is the chirality relative to this
counter-clockwise coordinate.

\subsection*{Drives, projection and calibration}

We compare three drives: modulated couplings of the bonds that touch the outer sites (as a gate on the couplers would
produce), couplings with a smooth profile of width $1.5$ lattice spacings (as from strain), and a modulated three-spin term (as
from a field). At low energy a local coupling $o_b$ acts on the edge as a constant plus the stress tensor at $x_b$, plus
operators of higher scaling dimension whose effect is smaller by powers of $Q_n$ times the lattice spacing. For a single
chiral Majorana there is no other operator of lower dimension that a bond coupling can produce (Supplementary
Note~\ref{si:sumrule}). The sum over bonds then picks one Fourier mode~\cite{KooSaleur1994}:
\begin{equation}
\begin{aligned}
O_n&\;\to\;a\,\frac{2\pi}{L}\,L_{-n}\qquad(n\neq0),\\
O_0&\;\to\;a\,\frac{2\pi}{L}\Big(L_0-\frac{c}{24}\Big)+\text{const},
\end{aligned}
\label{eq:projection}
\end{equation}
up to corrections of order $(n/L)^2$. The same coefficient $a$ appears in both, because both come from the same local
coupling.

The drive is absorbed at the frequencies of the transitions it causes. Fermi's golden rule gives the absorbed weight as the
sum of $|\langle j|O_n|i\rangle|^2$ over the transitions $i\to j$, weighted by the occupation difference $P_i-P_j$ of the two
states. We collect the weight of all transitions below the bulk gap and subtract the weight of the counter-propagating drive
$O_{-n}$. The drive acts on one edge only, so an opposite edge, which would absorb the other direction, is not excited. With
Eq.~\eqref{eq:projection}, the identity $\sum_{ij}(P_i-P_j)|\langle j|L_{-n}|i\rangle|^2=\langle[L_n,L_{-n}]\rangle$ and
Eq.~\eqref{eq:virasoro} with $m=n$,
\begin{equation}
\begin{aligned}
W_n&=a^2\Big(\frac{2\pi}{L}\Big)^2\langle[L_n,L_{-n}]\rangle\\
&=a^2\Big(\frac{2\pi}{L}\Big)^2\Big[2n\langle L_0\rangle+\frac{c}{12}(n^3-n)\Big].
\end{aligned}
\label{eq:Wn}
\end{equation}
The expectation value is taken in the state of the edge, whatever it is: pure or mixed, stationary or not (Supplementary
Note~\ref{si:sumrule}). The unknown prefactor $a^2(2\pi/L)^2$ is measured by the static change $\lambda O_0$. By
Eq.~\eqref{eq:projection} it shifts $H_{\rm edge}$ by $\lambda a(2\pi/L)L_0$, so the level spacing changes at the rate
$\partial\omega_F/\partial\lambda=a\,2\pi/L\equiv\vp\,2\pi/L$. Dividing Eq.~\eqref{eq:Wn} by its square gives
Eq.~\eqref{eq:Dn}.

Each excited state is assigned to a harmonic by its absorption. We use two spectral windows: a line window around each
harmonic, and a full window that collects all edge transitions below the gap. Flakes use the line window, because on flakes the
full window also collects weak lattice harmonics generated at the corners; on cylinders the two agree. The ladder calibration
tests Eq.~\eqref{eq:Dn} directly and converges as $1/L^2$. The line calibration involves the same lattice states as the line
weight, so their common lattice corrections largely cancel in the ratio, and in the systems we tested it converges much faster.
In the exact calculations we evaluate it as the weighted mean of the Hellmann--Feynman shifts of the states in the line, with
the weights held fixed. On flakes this agrees within $0.05\%$ with the shift of the exact line centroid at finite offsets
$\pm0.1$, which also includes the change of the weights; on cylinders the two agree within $2\%$. We use the calibration of the
$n=2$ line as the common denominator for all $D_n$. No value of $c$ enters the calibration.

\subsection*{Gaussian states}

For free Majoranas, $H=\frac14c^TMc$ with $M=iA$. A quadratic drive $O=\frac14c^TBc$, with $B=i\,\delta A$ antisymmetric, is
fully described by the matrix $B$. Any Gaussian state is fully described by $G=\langle cc^T\rangle$; for a thermal state
$G=2f(-M)$ with the Fermi function $f$. In the eigenbasis of $M$ (eigenvalues $\epsilon_p$), the weight of
$F(t)=\langle[O^\dagger(t),O]\rangle$ in a frequency window $\Omega$ is
\[
\begin{aligned}
W_\Omega=-\tfrac14\Big[&\mathrm{Tr}\big((B^\dagger\circ\Theta)BG\big)\\
&-\mathrm{Tr}\big(B(B^\dagger\circ\Theta)G\big)\Big],
\end{aligned}
\]
where $\circ$ is the element-wise product and $\Theta_{pq}=1$ if $\epsilon_q-\epsilon_p\in\Omega$ and $0$ otherwise. For
stationary states it reduces to the golden-rule sum of Supplementary Note~\ref{si:sumrule}. A kick $e^{-i\theta O_{\rm k}}$
with a quadratic $O_{\rm k}$ maps $G\to RGR^T$ with an orthogonal $R$; free evolution for a time $t$ maps
$G\to e^{At}Ge^{-At}$. These relations give all results on edge states, disorder and readout exactly.

\subsection*{Interacting edge Majoranas}

The interaction couples the two parallel bonds of every plaquette:
\begin{equation}
H=\frac{i}{4}\sum_{jk}A_{jk}c_jc_k+g\sum_p\sum_{a=x,y,z}(iu_bc_{b_1}c_{b_2})(iu_{b'}c_{b'_1}c_{b'_2}).
\label{eq:Hint}
\end{equation}
Here $b$ and $b'$ are the two bonds of type $a$ in plaquette $p$, and $A$ is the rim-closed flake described above. We pair
neighbouring Majoranas into complex fermions, $f_m=(c_{2m}+ic_{2m+1})/2$. Equation~\eqref{eq:Hint} then becomes an interacting
chain of $N/2$ spinless fermions with long-range couplings. We use two-site DMRG in TeNPy~\cite{TeNPy}, with fermion parity
conserved and $\chi\le512$. From the ground state and the next two states of the same parity we obtain the line weights
$|\langle k|O_{\pm n}|0\rangle|^2$ and the line calibration
$\partial_\lambda\omega=\langle k|O_0|k\rangle-\langle0|O_0|0\rangle$ (Hellmann--Feynman, no additional runs). In the vacuum
$D_1=0$, so $\Cest=2D_2$.

We order the Majoranas row by row and pair positions $2m$ and $2m+1$, so that $c_{2m}=f_m+f_m^\dagger$ and
$c_{2m+1}=i(f_m^\dagger-f_m)$. Each product of Majoranas in Eq.~\eqref{eq:Hint} and in the drives becomes a sum of products of
$f$ and $f^\dagger$ on at most four sites. TeNPy builds the matrix-product operator and inserts the Jordan--Wigner strings that
keep the fermionic signs. The gauge field is fixed in the vortex-free sector, so the matrix-product states carry only the
entanglement of the $c$ Majoranas. We run two-site DMRG with a subspace expansion, bond dimension increased in steps to $\chi$,
fermion parity conserved, and up to 24 sweeps. Excited states are found by DMRG restricted to the space orthogonal to the lower
states. Matrix elements $\langle k|O|0\rangle$ are exact contractions of the matrix-product states with the matrix-product
operator of the drive.

A DMRG of the spin model itself is not practical at these sizes. The $\mathbb Z_2$ gauge field adds a flat spectrum of
entanglement across every cut of the matrix-product state, and already 54 spins leave a $5\%$ truncation error at $\chi=128$.
On 24 spins the spin model is still exact, and we use it to check the reduction. With the rim operators ($h_R=2$), the
spin-model ground state is vortex free ($\langle W_p\rangle=1$), all rim operators equal $+1$, and its energy equals the
free-Majorana value to $1.1\times10^{-7}$.

\subsection*{Readouts}

\paragraph*{Echo.} The processor prepares the edge state and applies a traveling pulse of the edge couplings,
$\delta J_b(t)=2\lambda\,e^{-t^2/2\sigma_t^2}\cos(Q_nx_b-\omega_ct)$, with the carrier $\omega_c$ at the line. A Gaussian pulse
$\lambda[f(t)O_n+\text{h.c.}]$, with $f(t)=e^{-t^2/2\sigma_t^2}e^{-i\omega_ct}$, excites the edge state $|j\rangle$ with
amplitude $-i\lambda\langle j|O_n|0\rangle\int dt\,f(t)e^{i\omega_jt}$ at first order. The integral is
$\sqrt{2\pi}\sigma_te^{-(\omega_j-\omega_c)^2\sigma_t^2/2}$. On resonance the excitation probability, which is one minus the
return probability, is therefore $P=2\pi\sigma_t^2\lambda^2\sum_j|\langle j|O_n|0\rangle|^2=2\pi\sigma_t^2\lambda^2W_n$. At
larger $\lambda$, $P/\lambda^2$ decreases as $D(\lambda)=D_0+\alpha\lambda^2$. Two amplitudes $\lambda_1<\lambda_2$ give
$D_0=(qD(\lambda_1)-D(\lambda_2))/(q-1)$ with $q=(\lambda_2/\lambda_1)^2$ (Richardson extrapolation). For a single line,
$\ln P(\omega_c)$ is a parabola in $\omega_c$. Three carriers $\omega_0$ and $\omega_0\pm\delta$, with $\delta=1/\sigma_t$, give
the line centre $\omega_{\rm line}=\omega_0+\delta\,(\ln P_+-\ln P_-)/[2(2\ln P_0-\ln P_+-\ln P_-)]$. Repeating this at static
offsets $\pm\delta\lambda$, here $\pm0.1$ of the edge couplings, gives $\partial\omega_{\rm line}/\partial\lambda$. Ten
settings are needed: two directions at two amplitudes, and three carriers at two offsets. The shot budget follows from binomial
statistics, $\delta P=\sqrt{P(1-P)/S}$ for $S$ shots, propagated through these formulas.

\paragraph*{Correlator.} Write $F(t)=\int\Sigma(\omega)e^{-i\omega t}d\omega$ in terms of the absorption spectrum
$\Sigma(\omega)$. The weight in a window $(0,\omega_c)$ is $\int_0^{\omega_c}\Sigma\,d\omega=\frac1{2\pi}\int K(t)F(t)\,dt$ with
$K(t)=\int_0^{\omega_c}e^{i\omega t}d\omega$. With samples up to $t_{\max}$ we multiply by a smooth taper (a Blackman window),
which suppresses leakage from the bulk. The first moment, with $\omega$ inside the integral for $K$, gives the line centre and
hence the calibration from runs at static offsets.

\subsection*{Quantum-simulator circuits and error model}

The circuits use $\kappa=0.3$ and Trotter steps of $0.2/J$, with the bond layers placed symmetrically around the three-spin
layer. As for the circuit of Ref.~\cite{Ali2025}, the edge that is measured is that of the Floquet Hamiltonian of the step,
and the prepared vacuum is its ground state. The default pulse has $\sigma_t=2.5$ and is truncated at $\pm2.5\sigma_t$. The
twelve circuits are the ten settings of the echo readout and the two $n=1$ drives; Fig.~\ref{fig:qsim}(c) lists them in the
order $n=2$ at the two amplitudes (absorbing, then opposite direction), $n=1$ in both directions, and the three carriers at the
two offsets. The construction, the checks and the matrix-product-state settings are given in Supplementary
Note~\ref{si:circuits}.

For the error model we write $q_\lambda=1-P_\lambda$ for the observed return probability of a setting. If every error that
occurs removes the state from the return channel with the same probability $1-F$ in all circuits, then $q_\lambda=F(1-P)$ and
the three $\lambda=0$ references give $q_0=F$, so that $P=(q_0-q_\lambda)/q_0=(P_\lambda-P_0)/(1-P_0)$ recovers the ideal
excitation probability. We take $F=(1-p)^{N_2}$ with $N_2=4.3\times10^4$ compiled two-qubit gates. The shots for a $5\%$ error
on $\Cest$ follow from binomial noise on all settings and references, propagated linearly through the estimator; a Monte Carlo
resampling agrees within $4\%$. The model leaves out single-qubit gates, idle errors and measurement errors. Errors that
commute with all fluxes, such as a bond operator, which is a single Majorana bilinear, and coherent or setting-dependent errors
are not removed by the reference.



\bibliographystyle{apsrev4-2}
\bibliography{ref}


\clearpage
\onecolumngrid
\section*{Extended Data}

\renewcommand{\figurename}{Extended Data Fig.}
\setcounter{figure}{0}
\renewcommand{\tablename}{Extended Data Table}
\renewcommand{\thetable}{\arabic{table}}
\setcounter{table}{0}

\begin{figure*}[h]
\centering
\includegraphics[width=\textwidth]{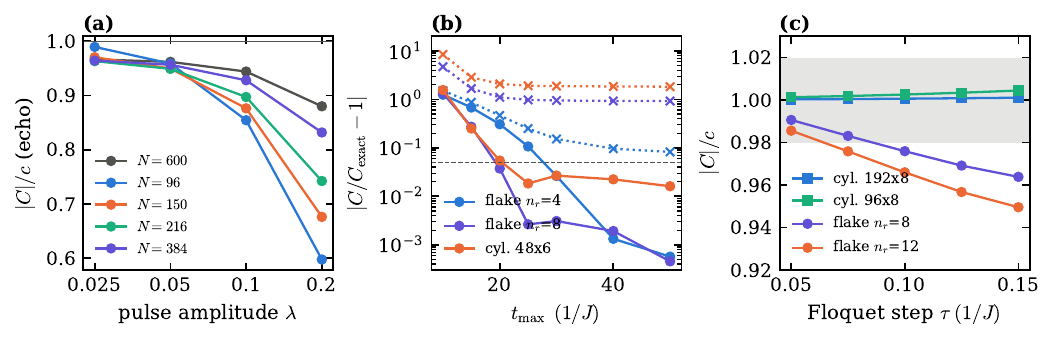}
\caption{Readout and Floquet diagnostics. The three panels are separate calculations. (a) Echo readout on flakes (without a
vison, $\sigma_t=4$): $|\Cest|/c$ from the exact excitation probabilities against the pulse amplitude, with the line
calibration from the same readout. (b) Correlator readout: error of $\Cest$ against the maximum evolution time, without
decoherence (solid) and with an uncorrected decay rate $\gamma=0.03$ (dotted). (c) The Floquet bond-layer circuit of
Ref.~\cite{Ali2025} with $V=0.3$: $|\Cest|/c$ (line calibration) against the Trotter step.}
\label{fig:protocol}
\end{figure*}

\begin{table*}[h]
\caption{$\Cest$ on the $L=512$, $W=8$ cylinder (7168 spins) for three drives, with the ladder (HF) and line (lc)
calibration, without (NS) and with (R) a flux through the cylinder. The last line gives $D_1^\R/2=h_\sigma$ (edge couplings,
ladder calibration); the exact value is $\frac1{16}=0.0625$.}
\label{tab:largest}
\begin{tabular}{llcccccc}
\toprule
 & & \multicolumn{2}{c}{edge couplings} & \multicolumn{2}{c}{smooth (strain)} & \multicolumn{2}{c}{three-spin}\\
$\kappa$ & sector & HF & lc & HF & lc & HF & lc\\
\midrule
0.1 & NS & 0.49986 & 0.50006 & 0.49969 & 0.49973 & 0.49986 & 0.49994\\
0.1 & R  & 0.49959 & 0.50007 & 0.49941 & 0.49952 & 0.49968 & 0.49988\\
0.2 & NS & 0.49996 & 0.50003 & 0.49987 & 0.49993 & 0.50001 & 0.50002\\
0.2 & R  & 0.49987 & 0.50003 & 0.49963 & 0.49979 & 0.49992 & 0.49996\\
0.3 & NS & 0.49999 & 0.50001 & 0.49994 & 0.49999 & 0.50009 & 0.50007\\
0.3 & R  & 0.49997 & 0.50002 & 0.49986 & 0.49996 & 0.50014 & 0.50009\\
\midrule
\multicolumn{8}{l}{$D_1^\R/2$: $0.06249$ ($\kappa=0.1$), $0.06250$ ($\kappa=0.2$), $0.06250$ ($\kappa=0.3$).}\\
\bottomrule
\end{tabular}
\end{table*}

\begin{table*}[h]
\caption{DMRG at the free point $g=0$ against the exact free-fermion values of the same flake (in parentheses where they
differ from the DMRG value in the shown digits): ground-energy error, first excitation $\omega_1$, line calibration and
$\Cest$.}
\label{tab:dmrg}
\begin{tabular}{rrccccc}
\toprule
$N$ & $\chi$ & $E_0$ error & $\omega_1$ (exact) & $\partial_\lambda\omega_F$ (exact) & $\Cest$ (exact) & relative error\\
\midrule
54 & 128 & $5.8\times10^{-6}$ & 1.65864 (1.65857) & 0.77132 (0.77134) & $-0.52167$ ($-0.52159$) & $1.6\times10^{-4}$\\
54 & 256 & $5.9\times10^{-8}$ & 1.65857 & 0.77134 & $-0.52159$ & $4\times10^{-6}$\\
96 & 128 & $2.9\times10^{-3}$ & 1.2748 (1.2657) & 0.53819 (0.53885) & $-0.50322$ ($-0.50546$) & $4.4\times10^{-3}$\\
96 & 256 & $1.9\times10^{-4}$ & 1.2667 & 0.53869 & $-0.50545$ & $1.4\times10^{-5}$\\
150 & 256 & $1.0\times10^{-2}$ & 1.0296 (1.0083) & 0.41252 (0.41306) & $-0.49614$ ($-0.50567$) & $1.9\times10^{-2}$\\
150 & 512 & $1.2\times10^{-3}$ & 1.0120 & 0.41278 & $-0.50452$ & $2.3\times10^{-3}$\\
\bottomrule
\end{tabular}
\end{table*}

\begin{table}[h]
\caption{Echo readout ($\kappa=0.3$, $\sigma_t=4$, pulse length $24/J$, amplitudes $\lambda=0.05,0.1$): $\Cest$ (without a vison,
NS), $h_\sigma$ (with a vison, R, from $D_1/2$), $\Lambda$, and the shots per setting for a $5\%$ standard error on $\Cest$. Ten
settings are needed: two directions at two amplitudes, and three carriers at two offsets.}
\label{tab:echo}
\begin{tabular}{lrcccc}
\toprule
system & $N$ & $\Cest$ & $h_\sigma$ & $\Lambda$ & shots (5\%)\\
\midrule
flake $n_r=4$ & 96 & $-0.497$ & $-0.074$ & $0.57$ & $2.0\times10^5$\\
flake $n_r=5$ & 150 & $-0.487$ & $-0.061$ & $0.507$ & $4.0\times10^5$\\
flake $n_r=5$, $\sigma_t=6$ & 150 & $-0.491$ & & & $1.3\times10^5$\\
flake $n_r=6$ & 216 & $-0.483$ & $-0.063$ & $0.518$ & $7.9\times10^5$\\
flake $n_r=8$ & 384 & $-0.483$ & $-0.061$ & $0.504$ & $2.5\times10^6$\\
flake $n_r=10$ & 600 & $-0.484$ & $-0.060$ & $0.494$ & $7.1\times10^6$\\
cylinder $24\times6$ & 240 & $0.493$ & $0.061$ & $0.504$ & $5.0\times10^5$\\
cylinder $32\times6$ & 320 & $0.490$ & $0.061$ & $0.501$ & $1.4\times10^6$\\
\bottomrule
\end{tabular}
\end{table}


\clearpage
\appendix
\renewcommand{\appendixname}{Supplementary Note}
\renewcommand{\thesection}{\arabic{section}}

\section{Derivation of the sum rule}\label{si:sumrule}

\paragraph*{Projection.} A bond coupling $o_b=-\sigma^\alpha_i\sigma^\alpha_j=iu_{ij}c_ic_j$ near the edge is bilinear in the $c$
Majoranas. At low energy the $c$ field near the edge is the chiral Majorana field $\psi(x)$. The bilinears of $\psi$ with the
lowest scaling dimension are the constant, and $\psi\partial_x\psi$, which is proportional to the stress tensor $T(x)$
(dimension 2)~\cite{DiFrancesco1997}. The term $\psi\psi$ is a constant for a Majorana field. All further terms have higher
dimension. A derivative term $\partial_xT$ adds $iQ_n$ times a constant to the coefficient $a$, which changes $|a|^2$ only at
order $Q_n^2$; terms such as $\partial_x^2T$ or $T^2$ are suppressed by $(Q_na_0)^2$, where $a_0$ is the lattice spacing.
Hence $o_b\to\epsilon_0+a\,T(x_b)\,\Delta x_b/(2\pi)+\dots$, with $\Delta x_b$ the edge length per bond. With the mode
expansion $T(x)=(2\pi/L)^2\sum_m(L_m-\frac{c}{24}\delta_{m0})e^{2\pi imx/L}$~\cite{DiFrancesco1997}, the sum in
Eq.~\eqref{eq:drive}, $\sum_be^{iQ_nx_b}\Delta x_b\to\int_0^Ldx\,e^{iQ_nx}$, selects $m=-n$ and gives
Eq.~\eqref{eq:projection}.

\paragraph*{Absorption.} A weak drive $\lambda(f(t)O_n+\text{h.c.})$ with a slowly varying envelope $f$ causes transitions
$i\to j$ with rate proportional to $|\langle j|O_n|i\rangle|^2$ at frequency $\omega_{ji}=E_j-E_i$. In a state with
occupations $P_i$, absorption and stimulated emission between the same two levels combine into the net weight
$(P_i-P_j)|\langle j|O_n|i\rangle|^2$. Summing over all pairs and using $O_n\propto L_{-n}$,
\[
\begin{aligned}
\sum_{ij}(P_i-P_j)|\langle j|L_{-n}|i\rangle|^2&=\sum_iP_i\langle i|L_nL_{-n}|i\rangle - \sum_jP_j\langle j|L_{-n}L_n|j\rangle\\
&=\langle[L_n,L_{-n}]\rangle .
\end{aligned}
\]
For a chiral edge, $L_{-n}$ raises the energy by $n\omega_F$, so all of this weight lies at positive frequency $n\omega_F$. The
counter-propagating drive $O_{-n}\propto L_n$ only lowers the edge energy and has no upward edge transitions. Subtracting its
upward weight removes contributions that are nearly the same for both directions, mainly bulk transitions. An edge of the
opposite chirality is not of this kind: it absorbs only $O_{-n}$. It does not contribute here because the drive is localized on
one edge.

\paragraph*{Non-stationary states.} For a state that is not diagonal in energy, the windowed weight also contains coherences
between states $i$ and $i'$. The operator $[L_n,L_{-n}]$ commutes with $L_0$ and with the edge momentum. Coherences between
states of different energy therefore drop out of its expectation value, and the identity holds for any density matrix. We
verified this numerically: the result does not depend on the delay of the probe (Supplementary Note~\ref{si:states}).

\section{Microscopic benchmarks and finite-size calibration}\label{si:bench}

With the ladder calibration on the largest cylinders (Extended Data Table~\ref{tab:largest}), the errors at $\kappa=0.2$ are
below $0.03\%$. The approach to the limit goes as $1/L^2$. It is slower at $\kappa=0.1$, where the edge mode is wider
($\Cest=0.348$ at $L=48$ and $0.4999$ at $L=512$).

The line calibration gives $|\Cest|=0.5216$, $0.5055$, $0.5057$, $0.5034$, $0.5021$ and $0.5009$ on flakes of $54$, $96$, $150$,
$216$, $384$ and $864$ spins ($\kappa=0.3$), and $0.5144$, $0.5061$ and $0.5034$ on $16\times6$, $24\times6$ and $32\times6$
cylinders. A 96-spin flake returns $c$ within $1.1\%$, and a 54-spin flake within $4.3\%$. At $\kappa=0.2$ the corresponding
sizes are 96 spins for $5\%$ and 600 spins for $1\%$. The sector ratio $\Lambda$ converges to $\frac12$ at the same rate: it is
$0.5046$ at 384 spins and $0.5001$ on the largest cylinders.

\section{Edge states, disorder and failure modes}\label{si:states}

\paragraph*{Thermal states.} We take the calibration once, in the ground state. $\Cest$ then stays within $1\%$ up to $T=0.1$
on $\kappa=0.3$ cylinders and up to $T=0.05$ to $0.075$ on flakes, while $D_1$ and $D_2$ grow by large factors. The deviation at
higher $T$ comes from the curvature of the edge dispersion, which the CFT neglects. It grows faster at $\kappa=0.2$
($T_{1\%}=0.025$), where the edge band is flatter.

\paragraph*{Non-stationary states.} We kick the ground state with traveling modulations of the edge couplings (one to three
harmonics $m\le3$) projected onto the states below the bulk gap. This is the ideal limit of a slow pulse, which does not excite
the bulk. The resulting state is Gaussian, and we evaluate the windowed absorption in it exactly. On the $96\times8$ cylinder,
all 15 states with $T_{\rm eff}\le T_{1\%}$ give $\Cest$ within $0.14\%$ of $c$. $\Cest$ also does not change when the probe is
delayed by $10$, $37$ or $100/J$ (spread $7\times10^{-4}$), although the state is not stationary. Sudden two-qubit gates on edge
bonds behave differently. A single gate of angle $0.05$ excites $0.003$ bulk Majorana quasiparticles. It shifts $\Cest$ by $4\%$
when the probe time is averaged, and by $20\%$ or more for a probe applied at once, because the gate creates coherences between
edge and bulk modes at first order.

\paragraph*{Vison disorder.} Random static visons leave $\Cest$ unchanged once no vison is within about five lattice spacings of
the edge [main Fig.~\ref{fig:states}(b), density $0.05$]. In separate scans at density $0.02$, for a clearance of at least 5, at
least $75\%$ of the configurations on cylinders and flakes are within $1\%$, and $94\%$ are within $5\%$ (median $1.0008$ to
$1.0020$). At density $0.1$ up to half of the configurations with large clearance still deviate by more than $1\%$. We
attribute this to the low-energy band formed by the Majorana modes bound to the visons.

\paragraph*{Bulk quasiparticles.} With a single occupied bulk Majorana mode, the median ratio of $\Cest$ to its clean value
lies between $-12$ and $7$ on cylinders of 672 and 1344 spins and flakes of 384 and 864 spins, because transitions within the
bulk band fall below the gap and into the edge windows. Thermal bulk excitations are negligible below $T\approx0.2$
($e^{-E_g/T}\lesssim5\times10^{-5}$ per mode).

\paragraph*{Miscalibrated couplings.} Independent random errors on the couplings bias the ladder calibration by
$\langle\vp^2\rangle/v_{\rm eff}^2-1$, the mismatch between the average of the local couplings and the effective edge velocity.
This bias reaches $3\%$ on the 384-spin flake at $\sigma_J=10\%$. On the 864-spin flake $\Cest/\Cest_{\rm clean}=0.9998$ at
$\sigma_J=5\%$, and $\Cest$ stays within $0.5\%$ of $c$ up to $\sigma_J=20\%$. Errors on the three-spin terms are smaller still.
We use flakes here: on cylinders strong disorder couples the two edges through accidentally degenerate levels.

\section{Interacting edge: benchmarks and scan}\label{si:interact}

\paragraph*{Benchmark.} Extended Data Table~\ref{tab:dmrg} compares DMRG with the exact solution at $g=0$. At $\chi=256$,
$\Cest$ agrees with the exact value to $4\times10^{-6}$ on 54 spins and $1.4\times10^{-5}$ on 96 spins. On 150 spins
$\chi=256$ is not converged ($\Cest$ is off by $1.9\%$). Doubling the bond dimension to $\chi=512$ reduces the error to $0.23\%$
and the energy error from $10^{-2}$ to $1.2\times10^{-3}$. The 150-spin interaction scan is run at $\chi=256$, and we show it as
a trend only.

\paragraph*{Interaction scan.} The scan covers $-0.4\le g\le0.4$. On 96 spins, $\omega_F$ changes from $+13\%$ to $-13\%$ of its
free value, and the raw line weight $W_2$ changes by up to $5\%$. The calibration follows the line weight. As a result,
$\Cest$ stays within $0.33\%$ of its free value for $|g|\le0.1$, within $0.7\%$ for $|g|\le0.2$, and within $1.5\%$ over the whole
scan. The remaining linear drift is a finite-size effect: at $|g|=0.2$ it is $1.7\%$, $0.7\%$ and $0.15\%$ on 54, 96 and 150
spins. The internal check $D_3/D_2$ rises from $3.50$ to $3.77$ to $3.88$ with size, towards the conformal value 4, and it does
not depend on $g$. The exception is 54 spins at $g=0.4$, where the $n=3$ line reaches the bulk continuum.

\section{Boundary field and Floquet circuit}\label{si:floquet}

\paragraph*{Boundary.} Reference~\cite{Ali2025} removes the degeneracy of the unpaired boundary Majoranas with a weak boundary
field $h\sigma^\alpha$ instead of the rim operators. The field couples a flat band of these Majoranas to the edge. On cylinders
this destroys the signal for every $h\le1$. On flakes the NS sector survives ($|\Cest|=0.500$ to $0.508$), but the R sector does
not: there is no signal for $h\le0.1$, and $D_1/2=-0.031$ to $-0.057$ instead of $-0.0625$ for $h=0.3$ to $3$. The boundary
field must therefore be avoided. Rim operators lift the degeneracy without coupling to the edge; they are three-qubit terms of
the same kind as the $\kappa$ term. A circuit that prepares the state directly needs no boundary term at all (Supplementary
Note~\ref{si:circuits}).

\paragraph*{Floquet circuit.} The trapped-ion implementation uses six layers of bond gates per step~\cite{Ali2025},
$U(\tau)=U_z(\tau-t_z)U_y(\tau-t_y)U_x(\tau-t_x)U_z(t_z)U_y(t_y)U_x(t_x)$, where $U_\alpha(t)$ applies all $\alpha$ bonds for a time
$t$. Its effective Hamiltonian contains a three-spin term of strength $V$. In the Majorana sector $U(\tau)$ maps Gaussian states
to Gaussian states, so we can compute its exact Floquet Hamiltonian. We test the circuit through the spectroscopy of this
Floquet Hamiltonian, separately from the echo simulations. The circuit edge obeys Eq.~\eqref{eq:C}: $\Cest=0.5004$ at Trotter
step $\tau=0.1$ on the $192\times8$ cylinder, and $0.5002$ to $0.5006$ for all $\tau\le0.15$ [Extended Data
Fig.~\ref{fig:protocol}(c)]. On flakes the corners add a $\tau$-dependent correction ($|\Cest|=0.483$ to $0.507$ at
$\tau=0.1$).

\section{Echo and correlator diagnostics}\label{si:readout}

\paragraph*{Echo.} We evolve the Gaussian state exactly, so the nonlinearity in $\lambda$ is included [Extended Data
Fig.~\ref{fig:protocol}(a)], and we extrapolate to $\lambda\to0$ from two amplitudes. The echo assumes a pure initial state;
our echo simulations start from the ground state. Extended Data Table~\ref{tab:echo} lists the results and the number of
measurements (shots) needed for a $5\%$ statistical error. The residual bias of about $3\%$ comes from the three-carrier
estimate of the line centre, which overestimates $\partial\omega_{\rm line}/\partial\lambda$ by $1.5\%$. A longer pulse
($\sigma_t=6$) reduces the bias to $0.7\%$ in the small-$\lambda$ limit and the shot count by a factor of three. From 150 spins
on, the echo also returns $h_\sigma$ within $4\%$, and $\Lambda$ within $1.4\%$ at 150 and 384 spins.

\paragraph*{Correlator.} A $5\%$ result requires a maximum evolution time $t_{\max}=15$ to $30/J$ [Extended Data
Fig.~\ref{fig:protocol}(b)], independent of the sampling step up to $dt=0.2$. A decay $e^{-\gamma t}$ of the signal, a simple
model of decoherence, broadens the bulk into the window. At $\gamma=0.03$ the uncorrected error is $8\%$ to $180\%$. Dividing
out the decay restores the exact result, and a $20\%$ error in the assumed rate leaves $1\%$ to $18\%$.

\section{Quantum-simulator circuits}\label{si:circuits}

\paragraph*{Layout and gates.} The physics patch has $5\times10$ grid sites, of which two corners are removed: 48 qubits, 63
bonds and 16 plaquettes. The number of columns is $2\bmod4$, so that the boundary sites pair up. The patch is placed where the
summed two-qubit and readout errors of a calibration snapshot are smallest. The gauge preparation starts from $|0\cdots0\rangle$
and maps single-qubit $Z$ stabilizers one by one onto the target stabilizers with gadgets $e^{i\pi Q/4}$, each a ladder of
controlled-NOT gates along a path of qubits. The targets are the plaquette operators, the bond operators of a reference
pairing of the $c$ Majoranas along a Hamiltonian path of bonds, and operators $\propto ib\,b'$ that pair the unpaired boundary
Majoranas consecutively along the same path. The vacuum is then reached from the reference pairing by nearest-neighbour Givens
rotations along the path, eliminating the target covariance from the far end~\cite{Langer2026}. Each rotation is a bond gate
$e^{i\phi\sigma^\alpha\sigma^\alpha}$, and the intermediate states are no more entangled than the target. A three-spin term is one
three-qubit rotation, compiled to four controlled-NOT gates. For the static offsets the vacuum of the offset couplings is
prepared. A gate that is a quadratic Majorana rotation $R$ maps the covariance $\Gamma$ to $R\Gamma R^T$, and the return
probability of two pure Gaussian states is $|\langle1|2\rangle|^2=|\det[(\Gamma_1+\Gamma_2)/2]|^{1/2}$. On 16 qubits a statevector
simulation confirms every stabilizer, the exact ground energy with an energy variance of $10^{-12}$, and the echo return
probabilities to $10^{-9}$, for both the gadget form and the three-qubit-unitary form of the three-spin gates.

\paragraph*{Compilation.} We compile one circuit for the couplers of the patch in the processor-native gate set (CZ and
single-qubit rotations) with Qiskit~\cite{Qiskit2024}. It contains $4.3\times10^4$ CZ gates. All bond and three-spin gates act
on couplers; 48 controlled-NOT gates in each gauge preparation and its inverse act on qubits that are not coupled and are routed
by the compiler. This is a resource estimate on the coupling graph, not a simulation of the calibrated device noise.

\paragraph*{Results on 48 qubits.} The twelve circuits give $\Cest=0.489$ from the exact excitation probabilities. Emulated
binomial shot noise with $10^6$ shots per setting gives $0.490\pm0.016$. Reversing $\kappa$ gives $\Cest=-0.492$. The two
magnitudes differ slightly, because neither the Trotter step, whose three-spin rotations are applied in a fixed order, nor the
discretized pulse is exactly symmetric under the reversal. For pulse widths $\sigma_t=2.5$, $3$ and $4$ and truncations at $2.5\sigma_t$, $3\sigma_t$ and
$4\sigma_t$, $\Cest$ lies between $0.481$ and $0.523$. The estimator assumes an untruncated Gaussian line shape, and this spread
is its finite-pulse systematic.

\paragraph*{Matrix-product states.} In the Majorana picture every Majorana pair across a cut carries half a bit, and the gauge
projection removes one bit. The Schmidt spectrum of the spin state across a cut of $L$ bonds with $k$ sites on one side is
therefore a flat factor $2^{(L+d+k\bmod2)/2-1}$, with $d\le1$ the number of boundary pairs across the cut, times the Gaussian
spectrum of the $c$ Majoranas. This reproduces the exact Schmidt spectra of 16-qubit states at every cut. Pairing the boundary
Majoranas along the matrix-product order keeps $d\le1$ and halves the required bond dimension at many cuts. A gate between
distant qubits can double the Schmidt rank across the cuts in between, and in our runs the working bond dimension is about
twice the static one. The static value is $\chi\approx60$ for the 28-qubit patch and $430$ for the 48-qubit patch at discarded
weight $10^{-6}$. We bring the qubits of every non-local gate together with nearest-neighbour SWAPs and undo them afterwards;
Aer would otherwise keep the permuted order, which scrambles the matrix-product state over a long circuit. A 28-qubit circuit
of about $8\times10^3$ gates then takes 15 to 30 minutes on one core at $\chi=256$. At $\chi=256$ the twelve excitation
probabilities agree with the exact values to $3\times10^{-6}$, and the resulting $\Cest$ of this patch agrees to four digits.
At $\chi=128$ the errors reach $10^{-3}$.

\paragraph*{Single errors.} On 16 qubits we insert random single-qubit Pauli errors into the vacuum preparation and the pulse
and compute the return probability with a statevector simulation. It stays below $10^{-30}$. Errors inside the Clifford gauge
preparation can return the state, and a bond-operator error in the pulse returns it with a probability of about one third.

\end{document}